\documentclass{nature-with-figures}

\usepackage{fancyhdr}
\usepackage[colorlinks=true,allcolors=blue]{hyperref}
\usepackage{amssymb}
\usepackage{amsmath}
\usepackage{graphicx}
\usepackage[caption=false]{subfig}
\usepackage{amsfonts}
\usepackage{xcolor}
\usepackage{epsfig}
\usepackage{color}
\usepackage{bm}
\usepackage{tabularx}
\usepackage{multirow}
\usepackage{mathtools}
\usepackage{xfrac}
\usepackage{blkarray}
\usepackage{bbold}
\usepackage[mathscr]{euscript}

\usepackage{comment}
\usepackage{authblk}

\newcommand{\ket}[1]{\vert{#1}\rangle} 
\newcommand{\bra}[1]{\langle{#1}\vert} 
 
\newcommand{\vnnb}{V$_{\text N}$N$_{\text B}$}

\title{Giant shift upon strain on the fluorescence spectrum of \vnnb~color centers in $h$-BN}

\author{Song Li$^1$, Jyh-Pin Chou$^1$,Alice Hu$^1$, Martin B. Plenio$^2$, P\'eter Udvarhelyi$^{3,4}$, Gerg\H{o} Thiering$^3$, Mehdi Abdi$^5$, \& Adam Gali$^{3,6}$}

\begin{document}

\maketitle

\begin{affiliations}
 \item Department of Mechanical Engineering, City University of Hong Kong, Hong Kong SAR, China
 \item Institute of Theoretical Physics and IQST, Albert-Einstein-Allee 11, Ulm University, 89069 Ulm, Germany
 \item Wigner Research Centre for Physics, P.O.\ Box 49, H-1525 Budapest, Hungary
 \item E\"otv\"os Science University, P\'azm\'any P\'eter S\'et\'any 1/A, H-1117 Budapest, Hungary
 \item Department of Physics, Isfahan University of Technology, Isfahan 84156-83111, Iran
 \item Budapest University of Technology and Economics,  Budafoki \'ut 8, H-1111 Budapest, Hungary
\end{affiliations}
\begin{addendum}
\item[Correspondence] Correspondence should be addressed to A.G.~(email: gali.adam@wigner.hu).
\end{addendum}
\newpage
\begin{abstract}
We study the effect of strain on the physical properties of the nitrogen antisite-vacancy pair in hexagonal boron nitride ($h$-BN), a color center that may be employed as a quantum bit in a two-dimensional material. With group theory and ab-initio analysis we show that strong electron-phonon coupling plays a key role in the optical activation of this color center. We find a giant shift on the zero-phonon-line (ZPL) emission of the nitrogen antisite-vacancy pair defect upon applying strain that is typical of $h$-BN samples. Our results provide a plausible explanation for the experimental observation of quantum emitters with similar optical properties but widely scattered ZPL wavelengths and the experimentally observed dependence of the ZPL on the strain.
\end{abstract}

%
%
%----------INTRODUCTION----------%
\section{Introduction}
Quantum emissions from two-dimensional (2D) materials has recently received considerable and rapidly rising interest of researchers in both condensed matter and quantum optics~\cite{Srivastava2015, Chakraborty2015, Palacios2016}
as these systems provide a potential basis for emerging technologies such as quantum nanophotonics~\cite{Xia2014, Clark2016, Shiue2017}, quantum sensing~\cite{Lee2012, Abderrahmane2014, Li2014}, and quantum information processing~\cite{Tran2016a, Abdi2017}.
The observation of single-photon emitters (SPEs) in hexagonal boron nitrite ($h$-BN) has added a fascinating new facet to the research field of layered materials~\cite{Aharonovich2016}.
The wide band gap of $h$-BN makes it an insulator that can host high quality emitters and 
allows for combination with other materials as substrates~\cite{Golberg2010}.
The experiments for exploring the nature of the emission of these color centers started with their first observations~\cite{Tran2016a, Tran2016b, Tran2016c, Jungwirth2016, Martinez2016, Schell2016, Shotan2016, Jungwirth2017, Li2017, Exarhos2017, Museur2008, Bourrellier2016, Vuong2016} and  recent theoretical works have provided evidence that the SPEs are indeed color centers, i.e. local point defects~\cite{Tawfik2017, Abdi2018, vanderWalle2018, vanderWalle2019, Alkauskas2019}.

Despite the considerable efforts that have been directed at the experimental exploration 
of these SPEs so far, a thorough theoretical understanding of the properties of the 
emitters that have been experimentally observed remains to be developed. Especially, the 
fact that many emitters appear at the edges of $h$-BN flakes and wrinkles on them~\cite{Chejanovsky2016, Ziegler2019} motivates the investigation on the effect 
of strain on these emitters. Furthermore, the quantum emitters have shown magnetic 
properties in some experiments~\cite{Toledo2018, Exarhos2019, Wrachtrup2019, Gottscholl2019}, while in other experiments non-magnetic behaviour was found~\cite{Li2017}. 

The most commonly observed quantum emitters exhibit emission in the visible, where competing theoretical models~\cite{Abdi2018,vandeWalle2019} exists in the literature. It has been suggested recently, based on the similarities of the optical lifetimes of the observed quantum emitters~\cite{Becher2019}, that two types of quantum emitters in the visible region may occur in $h$-BN, and the widely scattered zero-phonon-line (ZPL) energies might be attributed to external perturbations. One of the key candidates for this external perturbation is the local strain that may vary significantly in the $h$-BN samples, in particular, in polycrystalline $h$-BN samples. In another recent work, on the other hand, four different types of emitters have been found by means of combined chatodoluminescence and photoluminescence study where the applied stress did not change the brightness of emitters and the shift in ZPL was in the 10~meV region~\cite{Hayee2019}. Yet another experiment also found a relatively small shift upon applied stress on a given emitter~\cite{Grosso2017}.

These seemingly contradictory observations and the lack of a conclusive theoretical prediction motivates the study of the effect of strain on point defects in $h$-BN based on the assumption that point defects are the origin of the observed quantum emitters. Furthermore, a thorough understanding
of the electron-strain coupling properties also forms a rigorous theoretical basis for proposed spin- and electro-mechanical systems for control and manipulation of a mechanical resonator by means of spin-motion coupling~\cite{Abdi2017, Abdi2018a, Abdi2019}.
Group-theory analysis in combination with ab-initio Kohn-Sham density functional theory (DFT) simulations can be a very powerful tools for understanding the coupling between the optical emission and strain. So far the electron-strain coupling in $h$-BN emitters has been studied with limited accuracy~\cite{Grosso2017} by monitoring only the change in Kohn-Sham levels and the states that do not directly provide the ZPL energy in the PL spectrum monitored in the experiments.

Here we study the effect of strain on the ZPL emission of a key color center in $h$-BN, the nitrogen antisite-vacancy pair defect, by means of group theory and advanced DFT calculations, which can act as a quantum emitter with exhibiting a ZPL emission at around 1.9~eV (645~nm)~\cite{Abdi2018}. We report a giant, 12~eV/strain ZPL-strain coupling parameter for this quantum emitter which results in about 100~nm scattering of the ZPL emission with $\pm$1\% strain in $h$-BN sample. The physical origin of this giant effect is the strong electron-phonon coupling in $h$-BN. This result implies that local perturbations for vacancy type defects can seriously affect their optical spectrum and provide an explanation for the zoo of reported quantum emitters in the visible region. 

\begin{comment}
The rest of paper is organized as follows:
In the next section we provide a formulation for the effect of local strain on the energy levels of the color centers.
In Sec.~\ref{sec:} we investigate, by group theory analysis and \textit{ab initio} calculations, effect of strain on two color centers and in Sec.~\ref{sec:} magnetic properties are studied. The paper is closed by concluding remarks in Sec.~\ref{sec:}.
\end{comment}

\section{Results and Discussion}
\label{sec:results and Discussion}
\subsection{Group theory and DFT calculation analysis}
\begin{figure}
\includegraphics[width=\columnwidth]{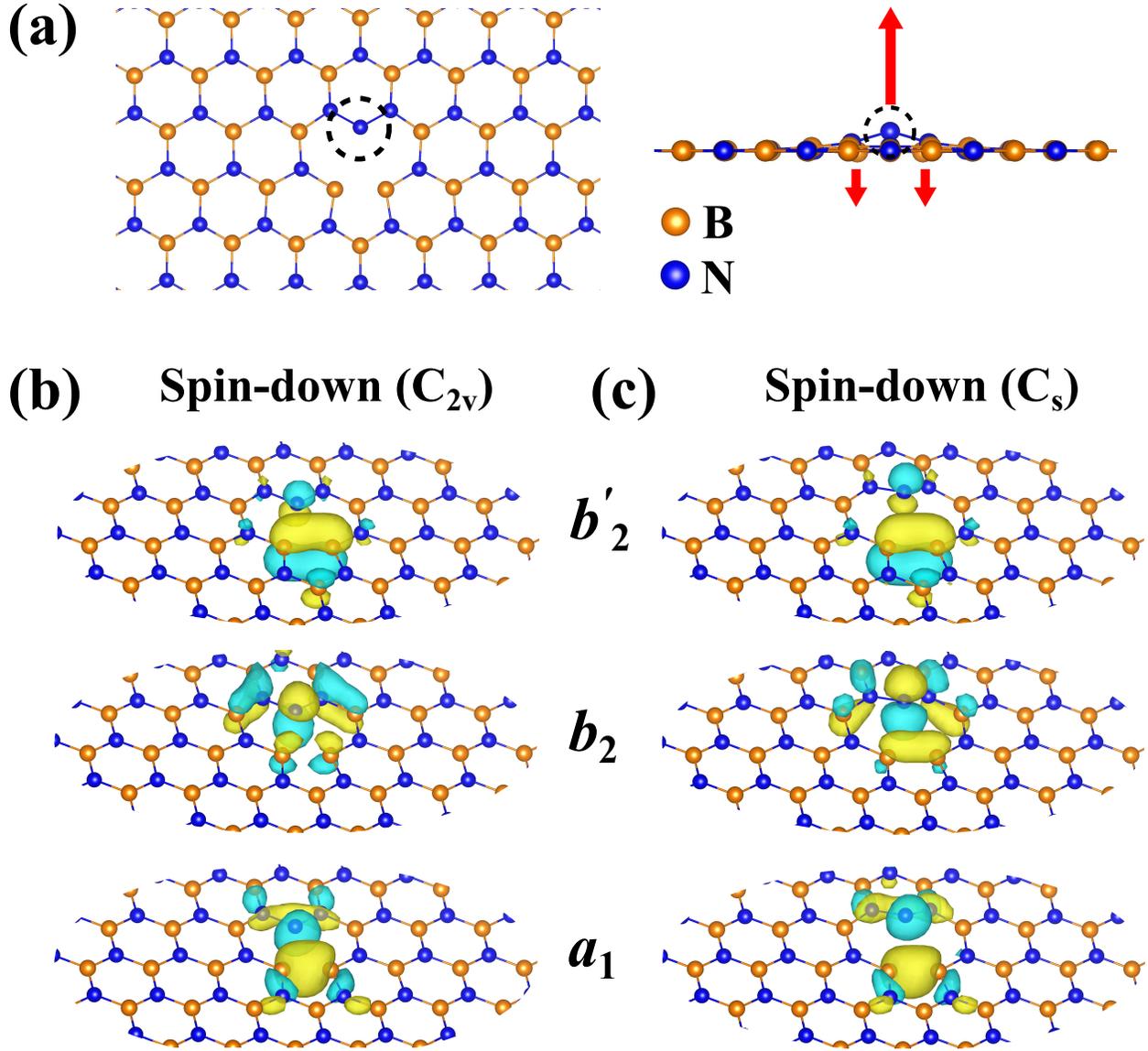}
\caption{\label{fig:Cs}%
Defect orbitals of \vnnb\ defect. (a) The geometry of \vnnb\ defect in the ground state. Top view (left) and side view (right). The dashed circle denotes the impurity nitrogen atom. The defect is out-of-plane and exhibits $\text{C}_\text{s}$ symmetry. The red arrows indicate the direction of the phonon vibration. (b) The $a_1$, $b_2$ and $b_2^\prime$ defect states in $\text{C}_\text{2v}$ geometry. (c) The defect states in the $\text{C}_\text{s}$ symmetry with the same energy order as those in (b). While in $\text{C}_\text{2v}$ symmetry only $b_2$ state has components out-of-plane, it can be seen that in $\text{C}_\text{s}$ symmetry the $a_1$ state also gains an out-of-plane component.}
\end{figure}

\begin{figure}
\includegraphics[width=\columnwidth]{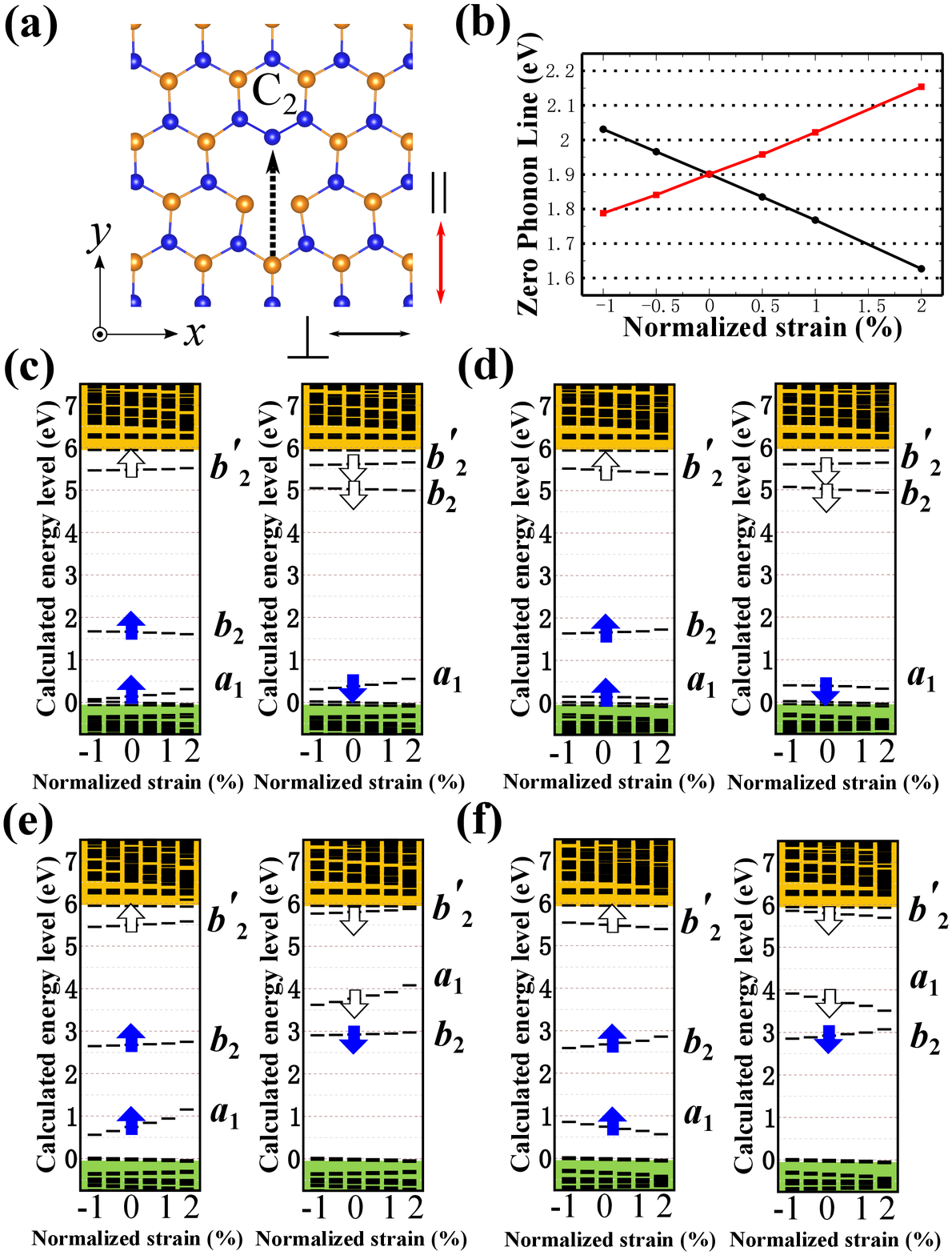}
\caption{\label{fig:strain}%
Zero-phonon-line energies upon strain for \vnnb\ defect. (a) The simplified cartoon of \vnnb\ defect in $h$-BN. The defect is not planar and exhibits $\text{C}_\text{s}$ symmetry in the ground state. The black and red arrow denote the directions of uniaxial strain perpendicular ($\perp$) and parallel ($\parallel$) to the ${\rm C_2}$ main axis, respectively. (b) The ZPL evolution as a function of external strain for $a_{1}$ to $b_{2}$ transition. The black and red lines denote the directions of strain perpendicular ($\perp$) and parallel ($\parallel$) directions. The energy level at ground state with $\text{C}_\text{s}$ symmetry evolution as a function of external strain for perpendicular (c) and parallel (d) directions, respectively. The energy level at excited state ($a_{1}$ to $b_{2}$) with $\text{C}_\text{2v}$ symmetry evolution as a function of external strain for perpendicular (e) and parallel (f) directions, respectively. The left and right panels show, respectively, the results for the spin-up channel and spin-down channels.}
\end{figure}

The sensitivity of the optical ZPL emission to the applied local strain greatly depends on the microscopic configuration of the point defect in  $h$-BN.
Here, we put our focus on the neutral nitrogen antisite-vacancy pair defect, \vnnb, which was first suggested as a candidate of the observed single-photon emissions~\cite{Tran2016a} and then thoroughly studied as one of the feasible quantum emitter in the visible with $S=1/2$ spin state~\cite{Abdi2018}. The optical properties of this defect have been examined from various theoretical methods and point of view~\cite{Abdi2018, Ali2018, Ping2019, Reimers2018}. 
The defect has a $\text{C}_\text{2v}$ symmetry before relaxation of the atoms in monolayer $h$-BN, and introduces three levels in the energy band gap that are labeled as $a_1$, $b_2$ and $b_2^\prime$ owing to their irreducible symmetry representation. These energy levels are occupied by three electrons resulting a $^{2}B_{2}$ many-electron ground state [see Figs.~\ref{fig:Cs} and \ref{fig:strain}]. In this paper, we use majuscule and minuscule to labels the symmetry of the many-elctron states and one-electron levels, respectively.
The two lowest-energy spin-preserving optical transitions are the following: in the spin majority channel the electron from $b_2$ may be promoted to $b_2^\prime$ or in the spin minority channel the electron from $a_1$ may be promoted to $b_2$. It is found that the $a_1 \leftrightarrow b_2$ has lower energy than $b_2 \leftrightarrow b_2^\prime$~\cite{Abdi2018, Ali2018, Ping2019}.
However, for the defect with the $\text{C}_\text{2v}$ symmetry the $a_1 \leftrightarrow b_2$ optical transition has a very small optical transition in-plane dipole moment~\cite{Ping2019}. Indeed, we also find this 
behavior in our own DFT calculation (see Supplementary Figure~S1).
We indeed notice that $b_2$ and $b_2^\prime$ states have wavefunctions that extend out-of-plane [see Fig.~\ref{fig:Cs}(b)], and therefore, can couple to phonon modes that drive the atoms out-of-plane, the membrane modes.
There is an unpaired electron in both $B_2$ ground state and $B_2^\prime$ excited state placed on the $b_2$ and $b_2^\prime$ orbital, respectively, that induce an out-of-plane geometry distortion. These phonon modes strongly couple the $^{2}B_{2}$ ground state and the $^{2}A_{1}$ excited state leading to the vibronic instablility of the ground state.
Therefore, the defect does not preserve the planar structure and the nitrogen antisite moves out from the plane in the ground state reducing the symmetry of the defect to $\text{C}_\text{s}$ [see Fig.~\ref{fig:Cs}(a)].
This geometry is about 100~meV lower in energy than the $\text{C}_\text{2v}$ configuration which reveals the strong coupling of the defect electrons to the membrane mode phonons.
This result basically agrees with previous DFT calculations~\cite{HosungNano2018}.
In $\text{C}_\text{s}$ symmetry, all the one-electron defect levels belong to $a^\prime$ irreducible representation. Despite the vibronic mixing, the correspondence to the high symmetry orbitals can be observed in Fig.~\ref{fig:Cs}(b) and (c).
When the hole is left at $a_1$ orbital in the $A_1$ excited state, then the coupling to the membrane phonons is negligible and the defects $\text{C}_\text{2v}$ symmetry is retrieved.
For the sake of simplicity, here we use the $\text{C}_\text{2v}$ symmetry labels for both configurations. The transformation of the point symmetry of the defect is discussed in Supplementary Note 1.

Owing to the rearrangement of the ions, the $B_2 (\text{C}_\text{s}) \leftrightarrow A_1 (\text{C}_\text{2v})$ optical transition assumes a dipole moment symmetry similar to that of
$B_2 (\text{C}_\text{s}) \leftrightarrow B_2^\prime (\text{C}_\text{s})$ transition. (see Supplementary Note 2)
This becomes clear from Fig.~\ref{fig:Cs}(c) where the wavefunctions $b_2$ and $a_1$ are shown for the $\text{C}_\text{s}$ configuration. In fact, both of them transform as $A^\prime$ in the $\text{C}_\text{s}$ configuration, and therefore, are coupled via the in-plane (the stronger component) polarization.
As a consequence, the lowest energy fluorescence is expected to occur as a radiative decay from the $A_1 (\text{C}_\text{2v})$ excited state to the $B_2 (\text{C}_\text{s})$ ground state because $A_1$ excited state has lower energy than that of $B_2^\prime$ excited state~\cite{Abdi2018}.
\emph{This result constitutes the optical activation of a color center in $h$-BN by means of a strong electron-phonon coupling.}
Consequently, it becomes important to study the strain dependence of the ZPL emission for the $B_2 (\text{C}_\text{s}) \leftrightarrow A_1 (\text{C}_\text{2v})$ optical transition. 
Details on the strain dependence for the higher energy transition is shown in the Supplementary Note~3.  

It is intriguing to carry out group theory analysis before starting the numerical ab-initio calculations. The detailed general description about the multi-electron configurations and their interaction with strain can be found in the Methods, that we apply to the \vnnb\ quantum emitter.
The analysis is performed for the $\text{C}_\text{2v}$ symmetry as the excited state transforms within $\text{C}_\text{2v}$ symmetry whereas the $\text{C}_\text{s}$ ground state follows the same analysis taking into account the fact that $\text{C}_\text{s}$ is a subgroup of the $\text{C}_\text{2v}$ symmetry.
Within $\text{C}_\text{2v}$ symmetry for the axial strain, the ZPL shift upon strain is given by 
\begin{equation}
\label{eq:shift}
\delta = \hat\varepsilon^{A_1}\big[\bra{b_2}\hat\Delta^{A_1}\ket{b_2} - \bra{a_1}\hat\Delta^{A_1}\ket{a_1}\big]\text{,}
\end{equation}
where $\hat\varepsilon^{A_1}$ is the strain tensor applying the strain parallel to the $C_2$ symmetry axis (axial strain) and $\hat\Delta^{A_1}$ is associated with the energy shifts for the corresponding electronic states (see Methods).
The $\text{C}_\text{s}$ configuration can then be described as an out-of-plane distortion due to a built-in strain which acts perpendicular to the basal plane. This mixes $a_1$ and $b_2$ orbitals through $B_2$ component of the strain as explained in Methods.
Since the energy spacing between the $a_1$ and $b_2$ levels is much larger than the typical deformation values this mixture should not significantly alter the energy shift of $\delta$ upon applying basal uniaxial strain. Hence, the energy shift in Eq.~\eqref{eq:shift} is a good approximation.
This ZPL energy shift is expected to depend linearly on strain for the $B_2 (\text{C}_\text{s}) \leftrightarrow A_1 (\text{C}_\text{2v})$ optical transition.
The magnitude of the strain-ZPL coupling as a function of the orientation of the applied uniaxial strain cannot be
determined by means of group theory. We, therefore, apply DFT simulations to quantify the strength of the strain-ZPL
coupling for \vnnb\ emission.

We calculate the ZPL energy as the total energy difference between the excited state and the ground state in the global energy minimum of the corresponding electronic configuration (see Method).
The strain is modelled by changing the lattice constant of the employed supercell [see Fig.~\ref{fig:strain}(a), where the parallel (red) and perpendicular (black) components of the strain are depicted]. The corresponding curves for the ZPL shift are plotted in Fig.~\ref{fig:strain}(b) as a function of the applied strain. The calculated ZPL of \vnnb\ without strain is 1.90~eV for the $B_2 (\text{C}_\text{s}) \leftrightarrow A_1 (\text{C}_\text{2v})$ optical transition, which is very close to the observed photoemission at 1.95~eV of certain SPEs~\cite{Tran2016b,Martinez2016}.
As expected, the curves are quasi-linear within $[-1;+2]$\% strain region, where we use $-$ and $+$ for compressive and tensile strain, respectively.
For both strain directions, we obtain a giant, 12~eV/strain shift in the ZPL energy which results in a huge variation in the emission wavelength as a function of strain.
For this quantum emitter, the tensile axial (perpendicular) uniaxial strain causes blueshift (redshift) in the ZPL emission. The variation of the ZPL energy with strain is about 100 nm and we believe our proposed mechanism yields the most plausible explanation for the observed variation of ZPL energies for certain SPEs.

In order to further study these results, we plot the shift of the defect levels upon the applied strain in the corresponding 
ground and excited states in Fig.~\ref{fig:strain}(c)-(f). Upon applying tensile axial strain, the $a_1$ level in the 
ground state shifts down whereas the occupied $b_2$ level shifts up in the ground and excited electronic configurations,
respectively. This will result in a blueshift in the ZPL. Upon applying tensile perpendicular strain, the $a_1$ level  
shifts up steeply whereas the occupied $b_2$ level moderately shifts up in the ground and excited electronic configurations,
respectively. As a consequence, the two levels approach each other upon applying this strain which results in a redshift 
in the ZPL energy. % could be explained why?  
This is in agreement with the observations reported in Ref.~\cite{Grosso2017} where the ZPL energy shift of the studied emitters exhibit a linear dependence on the applied strain. Furthermore, the three different possible orientations of the defect axis and our results on the blueshift (redshift) of the emission line for armchair (zigzag) strain explains the experimentally observed behavior~\cite{Grosso2017}.
  
\begin{comment}
 THIS CAN GO TO SUPPLEMENTARY.
 The $b_{2}$ to $b_2'$ excitation shows more dramatic nonlinear evolution of the ZPL with increased strain and this could be related to the out-of-plane strain between the impurity atom and the planar atoms. The $A_{1}\leftrightarrow B_{2}$ transition always has lower energy compared to the $B_{2}\leftrightarrow B_2^\prime$, thus it is justified to focus on the $A_{1}\leftrightarrow B_{2}$ optical transition for understanding the emission. 
\end{comment}

%
%
%

\subsection{Role of membrane phonons}
\begin{figure}
    \includegraphics[width=\columnwidth]{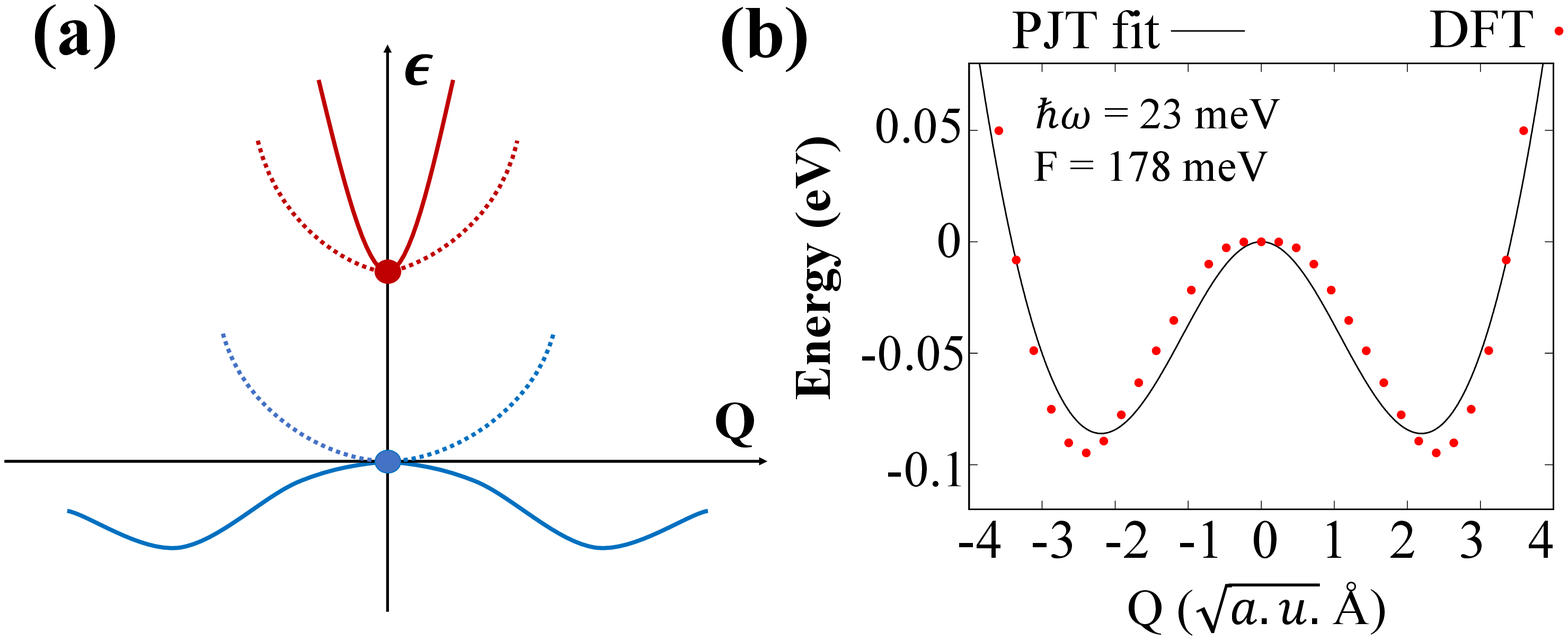}
    \caption{\label{fig:PJTE}Electron-phonon coupling of \vnnb\ defect. (a) Pseudo Jahn-Teller (PJT) effect between the excited state (red) and ground state (blue). Dashed line: before electron-phonon coupling; straight line: after electron-phonon coupling. $\epsilon$ is the total energy of the system whereas $Q$ is the selected configuration coordinate. (b) The calculated adiabatic potential energy surface of the ${}^2B_2$ ground state. The line is a fit of PJT model (see text), where the resultant values are $F$=178~meV and $\hbar \omega$=23~meV. The standard deviation is less than 2\%.}
\end{figure}

We find that the phonon modes with atoms moving out-of-plane, i.e., membrane modes, play a crucial role in the optical activation of the \vnnb\ defect in $h$-BN (see Supplementary Note~4). The modes are mainly contributed from the substitutional nitrogen atom. These membrane $B_2$ phonons couple the ${}^2A_{1}$ excited state and ${}^2B_{2}$ ground state where the ground state will be unstable at the $\text{C}_\text{2v}$ symmetry configuration and is distorted to $\text{C}_\text{s}$ configuration, whereas the excited state remains stable at $\text{C}_\text{2v}$ configuration. This suggest a strong pseudo Jahn-Teller (PJT)~\cite{Bersuker2006} system which is illustrated in Fig.~\ref{fig:PJTE}(a). 
We depict the adiabatic potential energy surface (APES) of the ${}^2B_2$ ground state of \vnnb\ in Fig.~\ref{fig:PJTE}(b) as obtained by HSE DFT calculations. We note that DFT calculation with a less accurate semilocal functional than HSE (see Method) obtained a similar APES in a previous study~\cite{HosungNano2018}. The Jahn-Teller energy is 95~meV. The solution of this strongly coupled electron-phonon system~\cite{Bersuker2006} is
\begin{equation}
\label{eq:PJTE}
\epsilon _{\pm} (Q) = \frac{1}{2}M\omega^2 Q^2 \pm \left(\Delta^2 + F^2Q^2\right)^\frac{1}{2} \text{,}
\end{equation}
where $Q$ is the normal coordinate of the effective phonon mode $\omega$ with the corresponding mass $M$, $\Delta$ is the energy gap between the ground state and excited state at the high symmetry point ($\text{C}_\text{2v}$ configurations) and $F$ is the strength of electron-phonon coupling. In the dimensionless generalized coordinate system we obtain $F$=178~meV and $\hbar \omega$=23~meV. This electron-phonon coupling parameter is about 2.5$\times$ larger than that of NV center in diamond~\cite{Thiering2017}. This indicates a giant electron-phonon interaction for the vacancy type defects in $h$-BN. We solved the electron-phonon PJT system quantum mechanically and found that the jumping rate between the two minima is 8.4~kHz. This is a relatively slow rate where the optical Rabi-oscillation between the ground and excited states should be more than two orders of magnitude faster (see Supplementary Note~5). This means that the ground state of ${}^2B_2$ is a static PJT system, and the ground state indeed exhibits low $\text{C}_\text{s}$ symmetry.

We note here that strong electron-phonon interaction with membrane phonons play an important role in the activation of intersystem crossing process in boron-vacancy optically detected magnetic resonance center~\cite{Ivady2019}. This type of phonon modes can be found only in 2D solid state systems. These findings demonstrate that the membrane phonon modes are major actors in the magneto-optical properties of solid state defect quantum bits and single photon emitters.

\subsection{Comparison to known single photon emitters}

Many SPEs were reported in multilayer $h$-BN structures. The physics of the membrane phonons and their effects on the optical properties for \vnnb\ defect are mainly discussed here based on the results achieved in monolayer $h$-BN, which directly models the single sheet $h$-BN flakes and can provide a tentative insight to the top layer of multilayered $h$-BN structures. We extend our study to the bulk $h$-BN model (see Methods) which correponds to such \vnnb\ defects that are buried deep in the multilayer $h$-BN structures. We find that the physics of the membrane phonons is the same: PJT occurs in the presence of van der Waals interaction but it suppresses the Jahn-Teller energy to 50.5~meV. Nevertheless, the resulting electron-phonon coupling remains strong with $F=193$~meV (see Supplementary Note~6). The weak interlayer interaction has little effect on the quantum emission of \vnnb\ as the ZPL energies and the ZPL energy shifts upon strain change less than 0.01~eV compared to the results obtained in the monolayer $h$-BN. 

Most of the SPEs in $h$-BN were first observed at room temperature~\cite{Tran2016a, Tran2016b} which emit in the visible region with various wavelength. Based on the ZPL emission region and the contribution of the phonon sideband (PSB) to the total emission, the visible SPEs were categorized into two groups according to early measurements~\cite{Tran2016b}: Group-1 with ZPL energies at 1.8-2.2~eV and with significant PSB contribution; Group-2 with ZPL energies at 1.4-1.8~eV with small PSB contribution. Group-1 emitters often showed an asymmetry in the ZPL lineshape at room temperature that was attributed to electron-phonon effects~\cite{Tran2016a}. Recently, low-temperature observation challenged this idea where they could decompose the PL spectra into two emitter components that could naturally explain the asymmetry of the spectrum at elevated temperatures where the spectra of the two emitters cannot be resolved~\cite{Bommer2019}. They found that the two emitters have ZPL energies with about 15~nm apart but very similar PSBs and also with a large variation of the ZPL wavelength of these pairs between 600~nm (2.06~eV) and 720~nm (1.72~eV) where the variation was tentatively attributed to strain~\cite{Bommer2019}. Based on the calculated ZPL energy and strong electron-phonon coupling of the \vnnb\ defect, the \vnnb\ definitely belongs to the Group-1 emitters and the calculated strain dependence of the calculated ZPL energies can cover the range of the observed ZPL energies by assuming about $\pm2$\% strain in the $h$-BN lattice. Parallel to our study, nanobeam electron diffraction has been applied to 
correlate the emitter optical emission with the emitter's local in-plane strain and found that about $\pm2$\% local strain can appear in $h$-BN flakes~\cite{Hayee2019}. They have found SPEs at 630~nm and 705~nm ZPL energies that are most likely connected by strain~\cite{Hayee2019}. The \vnnb\ defect can produce this giant shift of ZPL wavelength upon about $1.5$\% strain which is not far from the experimental observations claiming about $1$\%~\cite{Hayee2019}. We note that unambiguous
identification of the SPEs require further work from experiments and theory. Nevertheless, our study shows by means of accurate DFT calculations that the optical response of \vnnb\ defect to strain is indeed very sensitive. Further theoretical studies might reveal other defects in $h$-BN with similar properties.

\subsection{Implications towards quantum information processing applications}
The strong electron-phonon interaction implies that the optical properties of vacancy type quantum emitters in $h$-BN can vary significantly with strain. Indeed, our DFT simulations show a giant shift in the ZPL emission at strain values that can appear in $h$-BN. Compared to the SPEs in three-dimensional (3D) materials such as diamond and SiC, the SPEs in 2D $h$-BN are not buried by high refractive index medium which makes collection efficiency of the emitted light much higher than that for 3D materials.
Integration of $h$-BN based SPEs with nanophotonic devices offers a promising path to engineer quantum gates and circuitry which are key building blocks of quantum information processing and our work provides crucial implications in this field. First, we report the activation of the forbidden optical transition due to the strong electron-phonon coupling. The electron-phonon coupling reduce the symmetry of the ground state of the defect and it is a static PJT system. Second, we find the giant shift of the ZPL spectrum with external strain and propose a particular explanation for the phenomena. Until now, the spectral shift on SPEs in $h$-BN observed in our work is the largest known so far. Third, our result provides an analysis method for similar quantum emitters with varying ZPL energies and emission intensities in $h$-BN. Strain might be a reason for the spectral broadening and can influence the optical contrast and quantum efficiency. This might be harnessed to use this quantum emitter for realizing stress detector at the nanoscale as well as nanomechanical devices for quantum technologies. 

In this paper we performed a thorough group theory analysis and density functional theory calculations on the effect of strain on nitrogen antisite-vacancy color center in $h$-BN. We find a very strong electron-phonon interaction that can activate photoluminescence, and is responsible for the giant ZPL shift upon applied strain.
The behavior of the strain-induced energy shift is correlated with the experimental observations further revealing their microscopic nature.

%---- METHODS ----%
%%%%%%%%%%%
\section{Methods} 
\label{sec:methods}

\subsection{Group theory analysis on strain}
In this section we provide a formulation that describes effect of local strain of the point defects. The derived Hamiltonian 
is general and is then applied to the \vnnb\ quantum emitter in $h$-BN.

When considering point defects as candidates of the color centers in $h$-BN, the local strain manifests itself in modifying the atomic distances, which in turn, leads to the modification of the molecular orbitals (MOs) around the defect.
Any change in the properties of MOs result-in a redistribution of the energy states in the band structure of the host solid. Therefore, the strain directly couples to the electronic degrees-of-freedom of the color center. The coupling strength is called deformation potential.

We derive the Hamiltonian of the defect under local strain in the following way:
We start by assuming that the color center is composed of $N_\mathrm{e}$ valance electrons that are mostly isolated from the rest of lattice and gather around $N_\mathrm{n}$ nuclei forming up the defect. Note that this is a fairly good assumption by looking at the MOs drawn by DFT belonging to the defect states~\cite{Abdi2018}.
The attractive Coulomb energy imposed from nuclei on the electrons is then given by
\begin{equation}
V_{N_\mathrm{e}} = \sum_{j=1}^{N_\mathrm{e}}\sum_{k=1}^{N_\mathrm{n}} V_{jk}(\bm{\mathrm{x}}_k,\bm{\mathrm{x}}_j)
\approx -\sum_j\sum_k \frac{Z_k' \mathrm{e}^2}{R_{jk}},
\end{equation}
where $R_{jk}=|\bm{\mathrm{x}}_k -\bm{\mathrm{x}}_j|$ with $\bm{\mathrm{x}}_k$ position of the nuclei, while $\bm{\mathrm{x}}_j$ denotes location of the $j$th electron. Here, $Z_k'$ is the effective atomic number (screened nuclear charge) of the ion.
%Let us assume, for simplicity, that $V_{jk}$ only depends on $|\bm{R}_k -\bm{r}_j|\equiv R_{jk}$, the distance between the ions.
The local strain displaces ions involved in the point defect $\bm{\mathrm{x}} \rightarrow\bm{\mathrm{x}} +\delta\bm{\mathrm{x}}$ and thus their Coulomb interaction.
In the first order of accuracy we get $V(|\bm{\mathrm{x}}-\bm{\mathrm{x}}|)\longrightarrow V(|\bm{\mathrm{x}} +\delta\bm{\mathrm{x}} -\bm{\mathrm{x}}|)\approx V(R) +\big[\bm\nabla V(R)\big]_0\cdot\delta \bm{\mathrm{x}}$, where $\delta \bm{\mathrm{x}}$ is the infinitesimal displacement of the nuclei imposed by the local strain.
The value of displacement is obtained by $\delta\bm{\mathrm{x}}={\bm{\mathrm{X}}}\cdot\hat\varepsilon$ with the strain tensor $\hat\varepsilon$.
The electron--strain interaction Hamiltonian is then sum over all such first order variational terms
\begin{equation}
H_{\rm str} = \sum_{j,k}\big[\bm\nabla V(R_{jk})\big]_0\cdot\hat\varepsilon\cdot{\bm{\mathrm{X}}}_k \equiv -\sum_{j}\sum_\alpha \hat\Delta_j^\alpha\hat\varepsilon^\alpha,	
\label{strain}
\end{equation}
where we have introduced $\hat\Delta_{j} = {\bm{\mathrm{x}}}_{j}{\bm{\Xi}}_{j}$, a dyadic whose components have different group symmetries denoted by $\alpha$, while $\bm{\Xi}_{j} = \sum_k [\frac{1}{R_{jk}}\frac{\partial V_{jk}}{\partial R_{jk}}]_0 \bm{\mathrm{x}}_k$ is the deformation potential.
Here, we have assumed that the radial component of the gradient is the dominant one and neglected an irrelevant constant term. The former is a valid assumption as the total Coulomb attraction of the ions is more or less a central force~\cite{Boldrin2011, Peng2012, Singh2013}.

\renewcommand{\arraystretch}{1.5}
\begin{table}
\caption{\label{tab:vnnb_khonsa} The configuration, symmetry, and spin multiplicity of the ground and two lowest excited states of neutral \vnnb.}
%\begin{ruledtabular}
\begin{tabular}{lcc}
configuration & label & symmetry \\
%\colrule\rule{0pt}{2.5ex}%
$[a_1]^2[b_2]^1[b_2']^0$ & $\prescript{2}{}{B}_2$ & $\text{C}_\text{s}$ \\
$[a_1]^1[b_2]^2[b_2']^0$ & $\prescript{2}{}{A}_1$ & $\text{C}_\text{2v}$ \\
$[a_1]^2[b_2]^0[b_2']^1$ & $\prescript{2}{}{B}_2'$ & $\text{C}_\text{s}$ \\
\end{tabular}
%\end{ruledtabular}
\end{table}
\renewcommand{\arraystretch}{1.0}
The electronic configuration of this defect are given in Table~\ref{tab:vnnb_khonsa}.
%These states are sketched in Fig.~\ref{fig:vnnb}.
Given the orbital symmetries of these states and the following table of symmetry for strain components the group theory can predict that the only non-zero irreducible representations of the $\hat\Delta$ when sandwiched between two single-electron orbitals are:
\begin{equation*}
\text{C}_\text{2v}:
\begin{tabular}{c|ccc}
$\hat \Delta^{A_1}$ & $a_1$ & $b_2$ & $b_2'$ \\ \hline
$a_1$ & $\times$ & 0 & 0 \\
$b_2$ & 0 & $\times$ & $\times$ \\
$b_2'$ & 0 & $\times$ & $\times$ \\	
\end{tabular}~
\begin{tabular}{c|ccc}
$\hat \Delta^{B_2}$ & $a_1$ & $b_2$ & $b_2'$ \\ \hline
$a_1$ & 0 & $\times$ & $\times$ \\
$b_2$ & $\times$ & 0 & 0 \\
$b_2'$ & $\times$ & 0 & 0 \\	
\end{tabular},~
\text{C}_\text{s}:
\begin{tabular}{c|ccc}
$\hat \Delta^{A'}$ & $a_1$ & $b_2$ & $b_2'$ \\ \hline
$a_1$ & $\times$ & $\times$ & $\times$ \\
$b_2$ & $\times$ & $\times$ & $\times$ \\
$b_2'$ & $\times$ & $\times$ & $\times$ \\	
\end{tabular}.
\end{equation*}

The effect of strain on multi-electron states is a non-equal shift in their energy levels imposed by the axial components of the strain as well as inducing an interaction between the states via the axial and non-axial strain components.
The amount of shift only depends on the electronic states and its relations for the ground and excited states are $\delta_0$, $\delta_1$, and $\delta_2$, respectively. The strain prompted inter-state interactions are much smaller than the energy difference between the levels, hence one neglects them in an adiabatic manner.
The explicit form of the energy shifts are
\begin{align*}
\delta_0 &= \hat\varepsilon^{A'}\big[2\bra{a_1}\hat\Delta^{A'}\ket{a_1} +\bra{b_2}\hat\Delta^{A'}\ket{b_2}\big], \\
\delta_1 &= \hat\varepsilon^{A_1}\big[2\bra{b_2}\hat\Delta^{A_1}\ket{b_2} + \bra{a_1}\hat\Delta^{A_1}\ket{a_1}\big], \\
\delta_2 &= \hat\varepsilon^{A'}\big[2\bra{a_1}\hat\Delta^{A'}\ket{a_1} +\bra{b_2'}\hat\Delta^{A'}\ket{b_2'}\big].
\end{align*}
In the main text we have adopted the approximation that $\hat\varepsilon^{A'}\langle\cdot|\hat{\Delta}^{A'}|\cdot\rangle\approx \hat\varepsilon^{A_1}\langle\cdot|\hat{\Delta}^{A_1}|\cdot\rangle$ which is reasonable owing to the fact that $\text{C}_\text{s}$ is a subgroup of $\text{C}_\text{2v}$ and that the molecular orbitals retain their form.

\subsection{Details on DFT calculations}
The calculations are performed based on the density functional theory (DFT) implemented in Vienna ab initio simulation package (VASP).~\cite{Kresse1,Kresse2} Projector augmented wave (PAW) is used to separate the valence electrons from the core part. The energy cutoff for the expansion of plane-wave basis set was set to 450~eV which is enough to provide accurate result. The screened hybrid density functional of Heyd, Scuseria, and Ernzerhof (HSE)~\cite{Heyd} is used to calculate to band gap and defect levels. Within this approach, the short-range exchange potential is described by mixing with part of nonlocal Hartree-Fock exchange and this also provides reasonable geometry optimization of dynamic Jahn-Teller system. The HSE hybrid functional with mixing parameter of 0.32 closely reproduces the experimental band gap at 5.9~eV. To apply the strain along the parallel and perpendicular directions to the ${\rm C_2}$ axis, a $9\times5\sqrt{3}$ supercell is constructed through changing the basis to the orthorhombic structure. The perfect supercell contains 160 atoms which is sufficient to avoid the defect-defect interaction, and the single $\Gamma$-point scheme is converged for the k-point sampling for the Brillouin zone. The coordinates of atoms are allowed to relax until the force is less than 0.01~eV/\AA . The excited state was calculated within $\Delta$SCF method~\cite{Gali2009} that we previously applied to point defects in $h$-BN too~\cite{Abdi2018}. For the bulk simulation, a periodic model containing two layers are used, where one perfect layer is placed above the defective layer. The optimized interlayer distance is 3.37~\AA\ with DFT-D3 method of Grimme~\cite{Grimme2010}.

\begin{addendum}
 \item [Data Availability] The data that support the findings of this study are available from the corresponding author upon reasonable request.
 \item [Code Availability]The codes that were used in this study are available upon request to the corresponding author.
 \item AG acknowledges the Hungarian NKFIH grant No.~KKP129866 of the 
National Excellence Program of Quantum-coherent materials project, 
the National Quantum Technology Program (Grant No. 2017-1.2.1-NKP-2017-00001), and the EU H2020 Quantum Technology Flagship project 
ASTERIQS (Grant No.\ 820394). MBP acknowledges support from the ERC Synergy grant BioQ (Grant No.\ 319130), the EU H2020 Quantum 
Technology Flagship project ASTERIQS (Grant No.\ 820394), the EU H2020 Project Hyperdiamond (Grant No.\ 667192) and the BMBF via NanoSpin and DiaPol.
MA acknowledges support by INSF (Grant No.\ 98005028).
\item [Author contribution] SL carried out the DFT calculations under the supervision of JPC, AH, and AG. MA developed the group theory analysis with MBP. PU and GT developed and applied the electron-phonon coupling theory on the defect under the supervision of AG. All authors contributed to the discussion and writing the manuscript. AG conceived and led the entire scientific project.
 \item[Competing Interests] The authors declare that there are no competing interests..
\end{addendum}

%
%
%----------REFERENCES----------%
\section{References} 
\label{sec:references}
\bibliographystyle{naturemag}
\bibliography{origin}

\begin{thebibliography}{10}
\expandafter\ifx\csname url\endcsname\relax
  \def\url#1{\texttt{#1}}\fi
\expandafter\ifx\csname urlprefix\endcsname\relax\def\urlprefix{URL }\fi
\providecommand{\bibinfo}[2]{#2}
\providecommand{\eprint}[2][]{\url{#2}}

\bibitem{Srivastava2015}
\bibinfo{author}{Srivastava, A.} \emph{et~al.}
\newblock \bibinfo{title}{Optically active quantum dots in monolayer wse2}.
\newblock \emph{\bibinfo{journal}{Nat. Nanotechnol.}}
  \textbf{\bibinfo{volume}{10}}, \bibinfo{pages}{491--496}
  (\bibinfo{year}{2015}).

\bibitem{Chakraborty2015}
\bibinfo{author}{Chakraborty, C.}, \bibinfo{author}{Kinnischtzke, L.},
  \bibinfo{author}{Goodfellow, K.~M.}, \bibinfo{author}{Beams, R.} \&
  \bibinfo{author}{Vamivakas, A.~N.}
\newblock \bibinfo{title}{Voltage-controlled quantum light from an atomically
  thin semiconductor}.
\newblock \emph{\bibinfo{journal}{Nat. Nanotechnol.}}
  \textbf{\bibinfo{volume}{10}}, \bibinfo{pages}{507--511}
  (\bibinfo{year}{2015}).

\bibitem{Palacios2016}
\bibinfo{author}{Palacios-Berraquero, C.} \emph{et~al.}
\newblock \bibinfo{title}{Atomically thin quantum light-emitting diodes}.
\newblock \emph{\bibinfo{journal}{Nat. Commun.}} \textbf{\bibinfo{volume}{7}},
  \bibinfo{pages}{12978} (\bibinfo{year}{2016}).

\bibitem{Xia2014}
\bibinfo{author}{Xia, F.}, \bibinfo{author}{Wang, H.}, \bibinfo{author}{Xiao,
  D.}, \bibinfo{author}{Dubey, M.} \& \bibinfo{author}{Ramasubramaniam, A.}
\newblock \bibinfo{title}{Two-dimensional material nanophotonics}.
\newblock \emph{\bibinfo{journal}{Nat. Photon.}} \textbf{\bibinfo{volume}{8}},
  \bibinfo{pages}{899} (\bibinfo{year}{2014}).

\bibitem{Clark2016}
\bibinfo{author}{Clark, G.} \emph{et~al.}
\newblock \bibinfo{title}{Single defect light-emitting diode in a van der waals
  heterostructure}.
\newblock \emph{\bibinfo{journal}{Nano Lett.}} \textbf{\bibinfo{volume}{16}},
  \bibinfo{pages}{3944--3948} (\bibinfo{year}{2016}).

\bibitem{Shiue2017}
\bibinfo{author}{Shiue, R.-J.} \emph{et~al.}
\newblock \bibinfo{title}{Active 2d materials for on-chip nanophotonics and
  quantum optics}.
\newblock \emph{\bibinfo{journal}{Nanophotonics}} \textbf{\bibinfo{volume}{6}},
  \bibinfo{pages}{1329-- 1342} (\bibinfo{year}{2017}).

\bibitem{Lee2012}
\bibinfo{author}{Lee, S.} \emph{et~al.}
\newblock \bibinfo{title}{Flexible organic solar cells composed of p3ht:pcbm
  using chemically doped graphene electrodes}.
\newblock \emph{\bibinfo{journal}{Nanotechnology}}
  \textbf{\bibinfo{volume}{23}}, \bibinfo{pages}{344013}
  (\bibinfo{year}{2012}).

\bibitem{Abderrahmane2014}
\bibinfo{author}{Abderrahmane, A.} \emph{et~al.}
\newblock \bibinfo{title}{High photosensitivity few-layered mose2 back-gated
  field-effect phototransistors}.
\newblock \emph{\bibinfo{journal}{Nanotechnology}}
  \textbf{\bibinfo{volume}{25}}, \bibinfo{pages}{365202}
  (\bibinfo{year}{2014}).

\bibitem{Li2014}
\bibinfo{author}{Li, X.}, \bibinfo{author}{Yin, J.}, \bibinfo{author}{Zhou, J.}
  \& \bibinfo{author}{Guo, W.}
\newblock \bibinfo{title}{Large area hexagonal boron nitride monolayer as
  efficient atomically thick insulating coating against friction and
  oxidation}.
\newblock \emph{\bibinfo{journal}{Nanotechnology}}
  \textbf{\bibinfo{volume}{25}}, \bibinfo{pages}{105701}
  (\bibinfo{year}{2014}).

\bibitem{Tran2016a}
\bibinfo{author}{Tran, T.~T.}, \bibinfo{author}{Bray, K.},
  \bibinfo{author}{Ford, M.~J.}, \bibinfo{author}{Toth, M.} \&
  \bibinfo{author}{Aharonovich, I.}
\newblock \bibinfo{title}{Quantum emission from hexagonal boron nitride
  monolayers}.
\newblock \emph{\bibinfo{journal}{Nat. Nanotechnol.}}
  \textbf{\bibinfo{volume}{11}}, \bibinfo{pages}{37} (\bibinfo{year}{2016}).

\bibitem{Abdi2017}
\bibinfo{author}{Abdi, M.}, \bibinfo{author}{Hwang, M.-J.},
  \bibinfo{author}{Aghtar, M.} \& \bibinfo{author}{Plenio, M.~B.}
\newblock \bibinfo{title}{Spin-mechanics with color centers in hexagonal boron
  nitride membranes}.
\newblock \emph{\bibinfo{journal}{Phys. Rev. Lett.}}
  \textbf{\bibinfo{volume}{119}}, \bibinfo{pages}{233602}
  (\bibinfo{year}{2017}).
\newblock
  \urlprefix\url{https://journals.aps.org/prl/abstract/10.1103/PhysRevLett.119.233602}.

\bibitem{Aharonovich2016}
\bibinfo{author}{Aharonovich, I.}, \bibinfo{author}{Englund, D.} \&
  \bibinfo{author}{Toth, M.}
\newblock \bibinfo{title}{Solid-state single-photon emitters}.
\newblock \emph{\bibinfo{journal}{Nat. Photon.}} \textbf{\bibinfo{volume}{10}},
  \bibinfo{pages}{631} (\bibinfo{year}{2016}).

\bibitem{Golberg2010}
\bibinfo{author}{Golberg, D.} \emph{et~al.}
\newblock \bibinfo{title}{Boron nitride nanotubes and nanosheets}.
\newblock \emph{\bibinfo{journal}{{ACS} Nano}} \textbf{\bibinfo{volume}{4}},
  \bibinfo{pages}{2979--2993} (\bibinfo{year}{2010}).

\bibitem{Tran2016b}
\bibinfo{author}{Tran, T.~T.} \emph{et~al.}
\newblock \bibinfo{title}{Robust multicolor single photon emission from point
  defects in hexagonal boron nitride}.
\newblock \emph{\bibinfo{journal}{{ACS} Nano}} \textbf{\bibinfo{volume}{10}},
  \bibinfo{pages}{7331} (\bibinfo{year}{2016}).

\bibitem{Tran2016c}
\bibinfo{author}{Tran, T.~T.} \emph{et~al.}
\newblock \bibinfo{title}{Quantum emission from defects in single-crystalline
  hexagonal boron nitride}.
\newblock \emph{\bibinfo{journal}{Phys. Rev. Applied}}
  \textbf{\bibinfo{volume}{5}}, \bibinfo{pages}{034005} (\bibinfo{year}{2016}).

\bibitem{Jungwirth2016}
\bibinfo{author}{Jungwirth, N.~R.} \emph{et~al.}
\newblock \bibinfo{title}{Temperature dependence of wavelength selectable
  zero-phonon emission from single defects in hexagonal boron nitride}.
\newblock \emph{\bibinfo{journal}{Nano Lett.}} \textbf{\bibinfo{volume}{16}},
  \bibinfo{pages}{6052} (\bibinfo{year}{2016}).

\bibitem{Martinez2016}
\bibinfo{author}{Martinez, L.~J.} \emph{et~al.}
\newblock \bibinfo{title}{Efficient single photon emission from a high-purity
  hexagonal boron nitride crystal}.
\newblock \emph{\bibinfo{journal}{Phys. Rev. B}} \textbf{\bibinfo{volume}{94}},
  \bibinfo{pages}{121405(R)} (\bibinfo{year}{2016}).

\bibitem{Schell2016}
\bibinfo{author}{Schell, A.~W.}, \bibinfo{author}{Tran, T.~T.},
  \bibinfo{author}{Takashima, H.}, \bibinfo{author}{Takeuchi, S.} \&
  \bibinfo{author}{Aharonovich, I.}
\newblock \bibinfo{title}{Non-linear excitation of quantum emitters in
  two-dimensional hexagonal boron nitride}.
\newblock \emph{\bibinfo{journal}{ACS Photonics}} \textbf{\bibinfo{volume}{4}},
  \bibinfo{pages}{761--767} (\bibinfo{year}{2017}).
\newblock
  \urlprefix\url{https://pubs.acs.org/doi/abs/10.1021/acsphotonics.7b00025}.

\bibitem{Shotan2016}
\bibinfo{author}{Shotan, Z.} \emph{et~al.}
\newblock \bibinfo{title}{Photoinduced modification of single-photon emitters
  in hexagonal boron nitride}.
\newblock \emph{\bibinfo{journal}{{ACS} Photonics}}
  \textbf{\bibinfo{volume}{3}}, \bibinfo{pages}{2490} (\bibinfo{year}{2016}).

\bibitem{Jungwirth2017}
\bibinfo{author}{Jungwirth, N.~R.} \& \bibinfo{author}{Fuchs, G.~D.}
\newblock \bibinfo{title}{Optical absorption and emission mechanisms of single
  defects in hexagonal boron nitride}.
\newblock \emph{\bibinfo{journal}{Phys. Rev. Lett.}}
  \textbf{\bibinfo{volume}{119}}, \bibinfo{pages}{057401}
  (\bibinfo{year}{2017}).
\newblock
  \urlprefix\url{https://journals.aps.org/prl/abstract/10.1103/PhysRevLett.119.057401}.

\bibitem{Li2017}
\bibinfo{author}{Li, X.} \emph{et~al.}
\newblock \bibinfo{title}{Nonmagnetic quantum emitters in boron nitride with
  ultranarrow and sideband-free emission spectra}.
\newblock \emph{\bibinfo{journal}{{ACS} Nano}} \textbf{\bibinfo{volume}{11}},
  \bibinfo{pages}{6652--6660} (\bibinfo{year}{2017}).

\bibitem{Exarhos2017}
\bibinfo{author}{Exarhos, A.~L.}, \bibinfo{author}{Hopper, D.~A.},
  \bibinfo{author}{Grote, R.~R.}, \bibinfo{author}{Alkauskas, A.} \&
  \bibinfo{author}{Bassett, L.~C.}
\newblock \bibinfo{title}{Optical signatures of quantum emitters in suspended
  hexagonal boron nitride}.
\newblock \emph{\bibinfo{journal}{{ACS} Nano}} \textbf{\bibinfo{volume}{11}},
  \bibinfo{pages}{3328} (\bibinfo{year}{2017}).

\bibitem{Museur2008}
\bibinfo{author}{Museur, L.}, \bibinfo{author}{Feldbach, E.} \&
  \bibinfo{author}{Kanaev, A.}
\newblock \bibinfo{title}{Defect-related photoluminescence of hexagonal boron
  nitride}.
\newblock \emph{\bibinfo{journal}{Phys. Rev. B}} \textbf{\bibinfo{volume}{78}},
  \bibinfo{pages}{155204} (\bibinfo{year}{2008}).

\bibitem{Bourrellier2016}
\bibinfo{author}{Bourrellier, R.} \emph{et~al.}
\newblock \bibinfo{title}{Bright uv single photon emission at point defects in
  h‐bn}.
\newblock \emph{\bibinfo{journal}{Nano Lett.}} \textbf{\bibinfo{volume}{16}},
  \bibinfo{pages}{4317} (\bibinfo{year}{2016}).

\bibitem{Vuong2016}
\bibinfo{author}{Vuong, T. Q.~P.} \emph{et~al.}
\newblock \bibinfo{title}{Phonon-photon mapping in a color center in hexagonal
  boron nitride}.
\newblock \emph{\bibinfo{journal}{Phys. Rev. Lett.}}
  \textbf{\bibinfo{volume}{117}}, \bibinfo{pages}{097402}
  (\bibinfo{year}{2016}).

\bibitem{Tawfik2017}
\bibinfo{author}{Tawfik, S.~A.} \emph{et~al.}
\newblock \bibinfo{title}{First-principles investigation of quantum emission
  from hbn defects}.
\newblock \emph{\bibinfo{journal}{Nanoscale}} \textbf{\bibinfo{volume}{9}},
  \bibinfo{pages}{13575} (\bibinfo{year}{2017}).
\newblock
  \urlprefix\url{https://pubs.rsc.org/en/content/articlelanding/2017/nr/c7nr04270a/unauth#!divAbstract}.

\bibitem{Abdi2018}
\bibinfo{author}{Abdi, M.}, \bibinfo{author}{Chou, J.-P.},
  \bibinfo{author}{Gali, A.} \& \bibinfo{author}{Plenio, M.~B.}
\newblock \bibinfo{title}{Color centers in hexagonal boron nitride monolayers:
  A group theory and ab initio analysis}.
\newblock \emph{\bibinfo{journal}{{ACS} Photonics}}
  \textbf{\bibinfo{volume}{5}}, \bibinfo{pages}{1967--1976}
  (\bibinfo{year}{2018}).

\bibitem{vanderWalle2018}
\bibinfo{author}{Weston, L.}, \bibinfo{author}{Wickramaratne, D.},
  \bibinfo{author}{Mackoit, M.}, \bibinfo{author}{Alkauskas, A.} \&
  \bibinfo{author}{Van~de Walle, C.~G.}
\newblock \bibinfo{title}{Native point defects and impurities in hexagonal
  boron nitride}.
\newblock \emph{\bibinfo{journal}{Phys. Rev. B}} \textbf{\bibinfo{volume}{97}},
  \bibinfo{pages}{214104} (\bibinfo{year}{2018}).
\newblock \urlprefix\url{https://link.aps.org/doi/10.1103/PhysRevB.97.214104}.

\bibitem{vanderWalle2019}
\bibinfo{author}{Turiansky, M.~E.}, \bibinfo{author}{Alkauskas, A.},
  \bibinfo{author}{Bassett, L.~C.} \& \bibinfo{author}{Van~de Walle, C.~G.}
\newblock \bibinfo{title}{Dangling bonds in hexagonal boron nitride as
  single-photon emitters}.
\newblock \emph{\bibinfo{journal}{Phys. Rev. Lett.}}
  \textbf{\bibinfo{volume}{123}}, \bibinfo{pages}{127401}
  (\bibinfo{year}{2019}).
\newblock
  \urlprefix\url{https://link.aps.org/doi/10.1103/PhysRevLett.123.127401}.

\bibitem{Alkauskas2019}
\bibinfo{author}{Mackoit-Sinkevičienė, M.}, \bibinfo{author}{Maciaszek, M.},
  \bibinfo{author}{Van~de Walle, C.~G.} \& \bibinfo{author}{Alkauskas, A.}
\newblock \bibinfo{title}{Carbon dimer defect as a source of the 4.1 ev
  luminescence in hexagonal boron nitride}.
\newblock \emph{\bibinfo{journal}{Appl. Phys. Lett.}}
  \textbf{\bibinfo{volume}{115}}, \bibinfo{pages}{212101}
  (\bibinfo{year}{2019}).
\newblock \urlprefix\url{https://doi.org/10.1063/1.5124153}.

\bibitem{Chejanovsky2016}
\bibinfo{author}{Chejanovsky, N.} \emph{et~al.}
\newblock \bibinfo{title}{Structural attributes and photodynamics of visible
  spectrum quantum emitters in hexagonal boron nitride}.
\newblock \emph{\bibinfo{journal}{Nano Lett.}} \textbf{\bibinfo{volume}{16}},
  \bibinfo{pages}{7037} (\bibinfo{year}{2016}).

\bibitem{Ziegler2019}
\bibinfo{author}{Ziegler, J.} \emph{et~al.}
\newblock \bibinfo{title}{Deterministic quantum emitter formation in hexagonal
  boron nitride via controlled edge creation}.
\newblock \emph{\bibinfo{journal}{Nano Lett.}} \textbf{\bibinfo{volume}{19}},
  \bibinfo{pages}{2121--2127} (\bibinfo{year}{2019}).

\bibitem{Toledo2018}
\bibinfo{author}{Toledo, J.~R.} \emph{et~al.}
\newblock \bibinfo{title}{Electron paramagnetic resonance signature of point
  defects in neutron-irradiated hexagonal boron nitride}.
\newblock \emph{\bibinfo{journal}{Phys. Rev. B}} \textbf{\bibinfo{volume}{98}},
  \bibinfo{pages}{155203} (\bibinfo{year}{2018}).

\bibitem{Exarhos2019}
\bibinfo{author}{Exarhos, A.~L.}, \bibinfo{author}{Hopper, D.~A.},
  \bibinfo{author}{Patel, R.~N.}, \bibinfo{author}{Doherty, M.~W.} \&
  \bibinfo{author}{Bassett, L.~C.}
\newblock \bibinfo{title}{Magnetic-field-dependent quantum emission in
  hexagonal boron nitride at room temperature}.
\newblock \emph{\bibinfo{journal}{Nat. Commun.}} \textbf{\bibinfo{volume}{10}},
  \bibinfo{pages}{222} (\bibinfo{year}{2019}).

\bibitem{Wrachtrup2019}
\bibinfo{author}{Chejanovsky, N.} \emph{et~al.}
\newblock \bibinfo{title}{Single spin resonance in a van der waals embedded
  paramagnetic defect}.
\newblock \emph{\bibinfo{journal}{arXiv:1906.05903}}
  \urlprefix\url{https://arxiv.org/abs/1906.05903}.

\bibitem{Gottscholl2019}
\bibinfo{author}{Gottscholl, A.} \emph{et~al.}
\newblock \bibinfo{title}{Room temperature initialisation and readout of
  intrinsic spin defects in a van der waals crystal}  (\bibinfo{year}{2019}).
\newblock \urlprefix\url{http://arxiv.org/abs/1906.03774v1}.
\newblock \eprint{1906.0377}.

\bibitem{vandeWalle2019}
\bibinfo{author}{Turiansky, M.~E.}, \bibinfo{author}{Alkauskas, A.},
  \bibinfo{author}{Bassett, L.~C.} \& \bibinfo{author}{Van~de Walle, C.~G.}
\newblock \bibinfo{title}{Dangling bonds in hexagonal boron nitride as
  single-photon emitters}.
\newblock \emph{\bibinfo{journal}{Phys. Rev. Lett.}}
  \textbf{\bibinfo{volume}{123}}, \bibinfo{pages}{127401}
  (\bibinfo{year}{2019}).

\bibitem{Becher2019}
\bibinfo{author}{Bommer, A.} \& \bibinfo{author}{Becher, C.}
\newblock \bibinfo{title}{New insights into nonclassical light emission from
  defects in multi-layer hexagonal boron nitride}.
\newblock \emph{\bibinfo{journal}{Nanophotonics}} \textbf{\bibinfo{volume}{8}},
  \bibinfo{pages}{2041--2048} (\bibinfo{year}{2019}).

\bibitem{Hayee2019}
\bibinfo{author}{Hayee, F.} \emph{et~al.}
\newblock \bibinfo{title}{Revealing multiple classes of stable quantum emitters
  in hexagonal boron nitride with correlated optical and electron microscopy}.
\newblock \emph{\bibinfo{journal}{Nat. Mater.}} \textbf{\bibinfo{volume}{19}},
  \bibinfo{pages}{534--539} (\bibinfo{year}{2020}).

\bibitem{Grosso2017}
\bibinfo{author}{Grosso, G.} \emph{et~al.}
\newblock \bibinfo{title}{Tunable and high-purity room temperature
  single-photon emission from atomic defects in hexagonal boron nitride}.
\newblock \emph{\bibinfo{journal}{Nat. Commun.}} \textbf{\bibinfo{volume}{8}},
  \bibinfo{pages}{705} (\bibinfo{year}{2017}).

\bibitem{Abdi2018a}
\bibinfo{author}{Abdi, M.} \& \bibinfo{author}{Plenio, M.~B.}
\newblock \bibinfo{title}{Analog quantum simulation of extremely sub-ohmic
  spin-boson models}.
\newblock \emph{\bibinfo{journal}{Phys. Rev. A}} \textbf{\bibinfo{volume}{98}},
  \bibinfo{pages}{040303(R)} (\bibinfo{year}{2018}).

\bibitem{Abdi2019}
\bibinfo{author}{Abdi, M.} \& \bibinfo{author}{Plenio, M.~B.}
\newblock \bibinfo{title}{Quantum effects in a mechanically modulated
  single-photon emitter}.
\newblock \emph{\bibinfo{journal}{Phys. Rev. Lett.}}
  \textbf{\bibinfo{volume}{122}}, \bibinfo{pages}{023602}
  (\bibinfo{year}{2019}).

\bibitem{Ali2018}
\bibinfo{author}{Sajid, A.}, \bibinfo{author}{Reimers, J.~R.} \&
  \bibinfo{author}{Ford, M.~J.}
\newblock \bibinfo{title}{Defect states in hexagonal boron nitride: Assignments
  of observed properties and prediction of properties relevant to quantum
  computation}.
\newblock \emph{\bibinfo{journal}{Phys. Rev. B}} \textbf{\bibinfo{volume}{97}},
  \bibinfo{pages}{064101} (\bibinfo{year}{2018}).
\newblock \urlprefix\url{https://link.aps.org/doi/10.1103/PhysRevB.97.064101}.

\bibitem{Ping2019}
\bibinfo{author}{Wu, F.}, \bibinfo{author}{Smart, T.~J.}, \bibinfo{author}{Xu,
  J.} \& \bibinfo{author}{Ping, Y.}
\newblock \bibinfo{title}{Carrier recombination mechanism at defects in wide
  band gap two-dimensional materials from first principles}.
\newblock \emph{\bibinfo{journal}{Phys. Rev. B}}
  \textbf{\bibinfo{volume}{100}}, \bibinfo{pages}{081407}
  (\bibinfo{year}{2019}).
\newblock \urlprefix\url{https://link.aps.org/doi/10.1103/PhysRevB.100.081407}.

\bibitem{Reimers2018}
\bibinfo{author}{Reimers, J.~R.}, \bibinfo{author}{Sajid, A.},
  \bibinfo{author}{Kobayashi, R.} \& \bibinfo{author}{Ford, M.~J.}
\newblock \bibinfo{title}{Understanding and calibrating
  density-functional-theory calculations describing the energy and spectroscopy
  of defect sites in hexagonal boron nitride}.
\newblock \emph{\bibinfo{journal}{J. Chem. Theory Comput.}}
  \textbf{\bibinfo{volume}{14}}, \bibinfo{pages}{1602--1613}
  (\bibinfo{year}{2018}).
\newblock \urlprefix\url{https://doi.org/10.1021/acs.jctc.7b01072}.

\bibitem{HosungNano2018}
\bibinfo{author}{Noh, G.} \emph{et~al.}
\newblock \bibinfo{title}{Stark tuning of single-photon emitters in hexagonal
  boron nitride}.
\newblock \emph{\bibinfo{journal}{Nano Lett.}} \textbf{\bibinfo{volume}{18}},
  \bibinfo{pages}{4710--4715} (\bibinfo{year}{2018}).

\bibitem{Bersuker2006}
\bibinfo{author}{Bersuker, I.}
\newblock \emph{\bibinfo{title}{The Jahn-Teller Effect}}
  (\bibinfo{publisher}{Cambridge University Press}, \bibinfo{year}{2006}).

\bibitem{Thiering2017}
\bibinfo{author}{Thiering, G.} \& \bibinfo{author}{Gali, A.}
\newblock \bibinfo{title}{Ab initio calculation of spin-orbit coupling for an
  nv center in diamond exhibiting dynamic jahn-teller effect}.
\newblock \emph{\bibinfo{journal}{Phys. Rev. B}} \textbf{\bibinfo{volume}{96}},
  \bibinfo{pages}{081115} (\bibinfo{year}{2017}).
\newblock \urlprefix\url{https://link.aps.org/doi/10.1103/PhysRevB.96.081115}.

\bibitem{Ivady2019}
\bibinfo{author}{Iv\'ady, V.} \emph{et~al.}
\newblock \bibinfo{title}{Ab initio theory of negatively charged boron vacancy
  qubit in {hBN}}.
\newblock \emph{\bibinfo{journal}{npj Comput. Mater.}}
  \textbf{\bibinfo{volume}{6}}, \bibinfo{pages}{41} (\bibinfo{year}{2020}).
\newblock
  \urlprefix\url{https://www.nature.com/articles/s41524-020-0305-x#citeas}.

\bibitem{Bommer2019}
\bibinfo{author}{Bommer, A.} \& \bibinfo{author}{Becher, C.}
\newblock \bibinfo{title}{New insights into nonclassical light emission from
  defects in multi-layer hexagonal boron nitride}.
\newblock \emph{\bibinfo{journal}{Nanophotonics}} \textbf{\bibinfo{volume}{8}},
  \bibinfo{pages}{2041--2048} (\bibinfo{year}{2019}).
\newblock \urlprefix\url{https://doi.org/10.1515/nanoph-2019-0123}.

\bibitem{Boldrin2011}
\bibinfo{author}{Boldrin, L.}, \bibinfo{author}{Scarpa, F.},
  \bibinfo{author}{Chowdhury, R.} \& \bibinfo{author}{Adhikari, S.}
\newblock \bibinfo{title}{Effective mechanical properties of hexagonal boron
  nitride nanosheets}.
\newblock \emph{\bibinfo{journal}{Nanotechnology}}
  \textbf{\bibinfo{volume}{22}}, \bibinfo{pages}{505702}
  (\bibinfo{year}{2011}).

\bibitem{Peng2012}
\bibinfo{author}{Peng, Q.}, \bibinfo{author}{Ji, W.} \& \bibinfo{author}{De,
  S.}
\newblock \bibinfo{title}{Mechanical properties of the hexagonal boron nitride
  monolayer: Ab initio study}.
\newblock \emph{\bibinfo{journal}{Comput. Mater. Sci.}}
  \textbf{\bibinfo{volume}{56}}, \bibinfo{pages}{11} (\bibinfo{year}{2012}).

\bibitem{Singh2013}
\bibinfo{author}{Singh, S.~K.}, \bibinfo{author}{Neek-Amal, M.},
  \bibinfo{author}{Costamagna, S.} \& \bibinfo{author}{Peeters, F.~M.}
\newblock \bibinfo{title}{Thermomechanical properties of a single hexagonal
  boron nitride sheet}.
\newblock \emph{\bibinfo{journal}{Phys. Rev. B}} \textbf{\bibinfo{volume}{87}},
  \bibinfo{pages}{184106} (\bibinfo{year}{2013}).

\bibitem{Kresse1}
\bibinfo{author}{Kresse, G.} \& \bibinfo{author}{Furthm\"uller, J.}
\newblock \bibinfo{title}{Efficient iterative schemes for ab initio
  total-energy calculations using a plane-wave basis set}.
\newblock \emph{\bibinfo{journal}{Phys. Rev. B}} \textbf{\bibinfo{volume}{54}},
  \bibinfo{pages}{11169--11186} (\bibinfo{year}{1996}).
\newblock \urlprefix\url{https://link.aps.org/doi/10.1103/PhysRevB.54.11169}.

\bibitem{Kresse2}
\bibinfo{author}{Kresse, G.} \& \bibinfo{author}{Furthm\"uller, J.}
\newblock \bibinfo{title}{Efficiency of ab-initio total energy calculations for
  metals and semiconductors using a plane-wave basis set}.
\newblock \emph{\bibinfo{journal}{Comput. Mater. Sci.}}
  \textbf{\bibinfo{volume}{16}}, \bibinfo{pages}{15--50}
  (\bibinfo{year}{1996}).
\newblock
  \urlprefix\url{http://www.sciencedirect.com/science/article/pii/0927025696000080}.

\bibitem{Heyd}
\bibinfo{author}{Heyd, J.}, \bibinfo{author}{Scuseria, G.~E.} \&
  \bibinfo{author}{Ernzerhof, M.}
\newblock \bibinfo{title}{Hybrid functionals based on a screened coulomb
  potential}.
\newblock \emph{\bibinfo{journal}{J. Chem. Phys.}}
  \textbf{\bibinfo{volume}{118}}, \bibinfo{pages}{8207--8215}
  (\bibinfo{year}{2003}).
\newblock \urlprefix\url{https://doi.org/10.1063/1.1564060}.

\bibitem{Gali2009}
\bibinfo{author}{Gali, A.}, \bibinfo{author}{Janz\'en, E.},
  \bibinfo{author}{De\'ak, P.}, \bibinfo{author}{Kresse, G.} \&
  \bibinfo{author}{Kaxiras, E.}
\newblock \bibinfo{title}{Theory of spin-conserving excitation of the
  $n\ensuremath{-}{V}^{\ensuremath{-}}$ center in diamond}.
\newblock \emph{\bibinfo{journal}{Phys. Rev. Lett.}}
  \textbf{\bibinfo{volume}{103}}, \bibinfo{pages}{186404}
  (\bibinfo{year}{2009}).
\newblock
  \urlprefix\url{https://link.aps.org/doi/10.1103/PhysRevLett.103.186404}.

\bibitem{Grimme2010}
\bibinfo{author}{Grimme, S.}, \bibinfo{author}{Antony, J.},
  \bibinfo{author}{Ehrlich, S.} \& \bibinfo{author}{Krieg, H.}
\newblock \bibinfo{title}{A consistent and accurate ab initio parametrization
  of density functional dispersion correction (dft-d) for the 94 elements
  h-pu}.
\newblock \emph{\bibinfo{journal}{J. Chem. Phys.}}
  \textbf{\bibinfo{volume}{132}}, \bibinfo{pages}{154104}
  (\bibinfo{year}{2010}).

\end{thebibliography}

\end{document}